\documentclass[twocolumn]{aastex62}

\newcommand\Alfven{Alfv\'en \,}
\newcommand{\be}{\begin{equation}}
\newcommand{\ee}{\end{equation}}
\newcommand{\mE}{\mathcal{E}}
\newcommand{\mL}{\mathcal{L}}

\usepackage{amsmath}
\usepackage{epsfig}
\usepackage{epstopdf}
\usepackage{graphicx}
\usepackage{appendix}
\usepackage{txfonts}

\submitjournal{ApJ}
\shorttitle{Toward a Full MHD Jet Model}
\shortauthors{Huang et al.}

\begin{document}
\title{Toward a Full MHD Jet Model of Spinning Black Holes--II: Kinematics and Application to the M87 Jet}

\author{Lei Huang}
\affiliation{Shanghai Astronomical Observatory, Chinese Academy of Sciences, 80 Nandan Road, Shanghai 200030, China}
\affiliation{Key Laboratory for Research in Galaxies and Cosmology, Chinese Academy of Sciences, Shanghai 200030, China}
\affiliation{University of Chinese Academy of Sciences, 19A Yuquanlu, Beijing 100049, China}
\email{muduri@shao.ac.cn}

\author{Zhen Pan}
\affiliation{Perimeter Institute for Theoretical Physics, 31 Caroline St N, Waterloo, Ontario, N2L2Y5, Canada}
\email{zpan@perimeterinstitute.ca}

\author{Cong Yu}
\affiliation{School of Physics and Astronomy, Sun Yat-sen University, Zhuhai 519082, China}
\affiliation{State Key Laboratory of Lunar and Planetary Sciences, Macau University of Science and Technology, Macau, China}
\email{yucong@mail.sysu.edu.cn}

\begin{abstract}
In this paper, we investigate the magnetohydrodynamical structure of a jet powered by a spinning black hole,
where electromagnetic fields and fluid motion are governed by the Grad-Shafranov equation
and the Bernoulli equation, respectively.
Assuming steady and axisymmetric jet structure, the global solution is uniquely determined
with prescribed plasma loading into the jet and the poloidal shape of the outmost magnetic field line.
We apply this model to the jet in the center of nearby radio galaxy M87, and we find it can naturally explain
the slow flow acceleration and the flow velocity stratification within $10^5$ gravitational radii from the central black hole. In particular, we find the extremal black hole spin is disfavored by the
flow velocity measurements, if the plasma loading to the jet is dominated by the electron/positron pair production at the jet base.
\end{abstract}

\keywords{magnetic fields --magnetohydrodynamics (MHD) --black hole physics --galaxies: jets --galaxies: individual (M87)}

\section{Introduction} \label{sec:intro}

Many open questions on the nature of the black hole (BH) jets still remain to be answered :  what is the central engine, how the fluid is loaded into the jet and is accelerated to relativistic speed, and how the jet is collimated? \citep[see e.g.,][for reviews]{Meier01, Blandford19}
The well-resolved nucleus of nearby elliptical galaxy M87, which powers a highly collimated jet,
provides the best natural laboratory to study the jet physics.
The shadow image of the super-massive BH candidate in the center of M87 has been obtained
by the Event Horizon Telescope (EHT) \citep{EHT19I,EHT19II,EHT19III,EHT19IV,EHT19V,EHT19VI},
which provides a direct evidence of the existence of a super-massive BH with mass $(6.5\pm0.7)\times10^9M_\odot$ and nonzero spin.
On a larger scale ($\lesssim 10^2$ pc from the central BH), both the shape and
the proper motion of the M87 jet have been extensively measured by different collaborations since decades ago,
e.g., Hubble Space Telescope \citep{Bire99}, Very Long Baseline Array \citep{Cheung07, Hada11,Hada13,Hada16, Mertens16,Park19},
European VLBI Network \citep{Asad14}, and the KVN and VERA Array \citep{Park19}.
As discovered by \cite{Asad12} and \cite{Hada16}, the radial profile of the jet width is well approximated
by a parabola $W_{\rm jet}(r)\propto r^{0.56\sim 0.69}$, while the proper motion shows much more complicated behaviour.
The proper motion is reported to cover a wide range, from subluminal ($<0.1 c$) to superluminal ($\sim3 c$),
at distances of $10^{-1}-10^2$ pc along the jet from BH.
For a even further spot, HST-1, an X-ray knot located at $\approx 1$ kpc from the central BH, \cite{Snio19} reported the fastest superluminal motion with apparent speed of $\approx 6.3 c$.

On the theory side, the most promising candidate powering the BH jet is
the Blandford-Znajek (BZ) mechanism \citep{BZ77}, in which an electromagnetic
(EM) process occurs to extract the rotation energy of the BH in the form of Poynting flux.
The BZ mechanism has been extensively examined against the M87 jet by utilizing general relativistic magnetohydrodynamic (GRMHD) simulations, as well as the steady axisymmetric force-free electrodynamic (FFE) solution. For example, recent GRMHD simulations done by \cite{Naka18} imply that the parabolic shape of the jet is a generic consequence of interaction with the matter-dominated wind. Limited by computation speed, simulations are usually limited to $\lesssim 10^2$ gravitational radii, therefore there is no way to examine the simulated jet structure on a larger scale. As a complement, FFE solutions \citep[see e.g.,][]{Tanabe08,Contop13,Nathan14,Gralla15,Gralla16,Pan14,Pan15,Pan15b,Pan16,Pan17,Pan18,
Yang14, Yang15, East18, Mahlmann18} has been used to explore the jet structure on larger scales,
while it carries no information of flow velocity within the jet. To circumvent this problem, \cite{NMF07} proposed to
use the drift velocity of charged particles in the EM field as a proxy of the flow velocity,
which is found to greatly overestimate the flow velocity of the M87 jet \citep{Naka18}.
On the other hand, the flow velocity on prescribed magnetic field lines has also been explored by many authors
\citep[see e.g.,][]{Fendt01,Fendt04, Globus13, Globus14,Pu12,Pu15,Pu20, Polko13,Polko14,Beskin98,Beskin17,Cecco18}.
As we will show in this paper later, the flow velocity is sensitive to the shape of the magnetic field lines, which
should be obtained from first principle instead of be prescribed \emph{a priori} for making trustworthy predictions.

\begin{figure}[h!]
\centering
\includegraphics[scale=1]{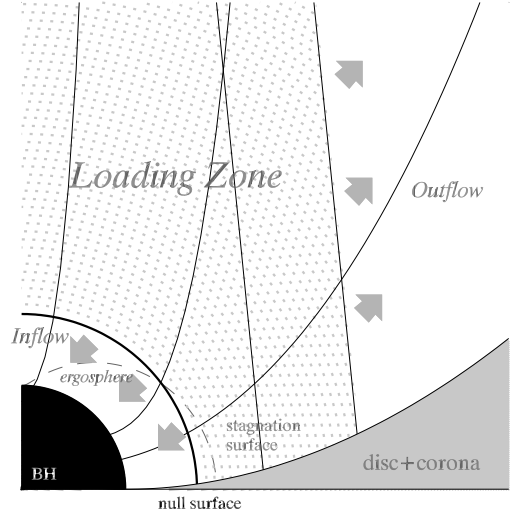}
\caption{A cartoon picture of the MHD model of BH jets.
The dotted-shaded region represents an extended plasma loading zone
where particles are injected into the jet (see Section~\ref{sec:GS} for more details).
The inner boundary of the loading zone is the null surface, shown in thick black line.
The stagnation surface, which locates inside the loading zone, is shown in thin line.
Both the inflow and outflow patterns form under the mutual influence of the BH and the EM fields naturally.
The gray shade represents the region fulfilled by disc and corona.
\label{fig:cartoon}}
\end{figure}

In \citet{Huang19} (hereafter Paper I), we constructed a GRMHD framework to investigate
the steady and axisymmetric structure of spinning BH powered jet (see Fig.~\ref{fig:cartoon} for a
cartoo picture of this jet model), where the fluid motion and the EM fields are self-consistently solved.
A pedagogical example of monopole jet was presented therein.
In this paper, we aim to apply this framework to collimated jets, especially to the M87 jet, and to explain
its kinematics which has not been well understood before.

This paper is organized as follows.
In  Section~\ref{sec:GS}, we summarize the basic assumption, introduce the Bernoulli equation, the MHD Grad-Shafranov (GS) equation,
and the singular surfaces of these two equations.
In Section~\ref{sec:results}, we present the numerical solution of a fiducial jet model
with super-magnetosonic inflow and outflow, explore more model parameter space and compare the model predictions
with the observations of the M87 jet, especially the radial profile of flow velocity.
Summary and discussion are given in Section~\ref{sec:summary}.
The numerical techniques for solving the governing equations are detailed in appendices.
We adopt the geometrical units $c=G=1$, and the mass/gravitational radius of the BH in the center of M87 $M=6.5\times10^9M_\odot \approx 3.2\times 10^{-4} $ pc throughout this paper.

\section{Governing Equations} \label{sec:GS}
Following Paper I, we assume a steady and axisymmetric jet ($\partial_t = \partial_\phi = 0$) in the Kerr spacetime with metric $ds^2 = g_{\mu\nu}dx^\mu dx^\nu$ with $\mu,\nu =\{t, r,\theta, \phi\}$,  and  perfectly conducting fluid within the jet, then EM fields $F_{\mu\nu}$ are completely determined by three functions:
the magnetic flux $\Psi(r,\theta)$, the angular velocity of magnetic field lines $\Omega(\Psi)$, and the electric current $I(r,\theta)$. We consider cold fluid within the jet, and we may define three quasi-conserved quantities:
the rest mass flux per magnetic flux $\eta_m$
\begin{equation}\label{eq:eta}
\begin{aligned}
 \eta_m(\Psi)\equiv\frac{\sqrt{-g} \rho u^r}{\Psi_{,\theta}} = - \frac{\sqrt{-g} \rho u^\theta}{\Psi_{,r}}
 = \frac{\sqrt{-g} \rho (u^\phi-\Omega u^t)}{F_{r\theta}} \ ,
\end{aligned}
\end{equation}
the  energy flux per magnetic flux $\mathcal{E}$
and the angular moment flux per magnetic flux $\mathcal{L}$
\begin{equation}\label{eq:EandL}
\begin{aligned}
	\mathcal{E}(\Psi)&:= -\eta_m u_t +\frac{\Omega I}{4\pi}\ ,  \\
	\mathcal{L}(\Psi)&:= \eta_m u_\phi +\frac{I}{4\pi}\ .
\end{aligned}
\end{equation}
where $\sqrt{-g}$ is the square root of the Kerr metric determinant, $\rho$ is the rest mass density and $u^\mu$ is the fluid four-velocity.
These three quantities are conserved only outside the plasma loading zone, i.e., $D_\Psi^\parallel X = 0$ is
true only where there is no plasma loading. Here $X = \{\eta_m ,\mE, \mL\}$ and
\begin{eqnarray}
	D^\parallel_\Psi&\equiv&\frac{1}{\sqrt{-g}}(\Psi_{,\theta}\partial_r-\Psi_{,r}\partial_{,\theta})\ ,
\end{eqnarray}
is the derivative along field lines.
Therefore there will be two sets of conserved quantities
$X_{\rm in}$ and $X_{\rm out}$
connecting to the inflow and the outflow, respectively.

\subsection{Bernoulli equation} \label{sec:Bern}

The relativistic Bernoulli equation is in fact the the normalization condition $u^\mu u_\mu=-1$
\begin{eqnarray}\label{eq:Bern}
	\mathcal{F}(u)&=&u_p^2+1-\left(\frac{\mE}{\eta_m}\right)^2 U_g(r,\theta, \mathcal M^2; \Omega, \mE/\mL)=0\ ,
\end{eqnarray}
where $u_p^2 \equiv u^Au_A$ with $A=\{r,\theta\}$.
$U_g$ is a function \citep{Camen86a,Camen86b,Camen87,Fendt01,Fendt04,Levinson06,Nitta91,Takahashi90,Pu15} of poloidal coordinates $r,\theta$, conserved quantities $\Omega, \mE/\mL$ and
the \Alfven Mach number $\mathcal{M}$ which is defined by
\begin{equation}\label{eq:mach}
    \mathcal{M}^2=\frac{4\pi \eta_m^2}{\rho} = 4\pi \eta_m \frac{u_p}{B_p},
\end{equation}
with $(\sqrt{-g} B_p)^2 = g_{rr} (\Psi_{,\theta})^2 + g_{\theta\theta} (\Psi_{,r})^2$
defining the poloidal magnetic field $B_p$.

\subsection{MHD GS equation} 

The MHD GS equation is  written in a compact form (see Paper I for detailed derivation)
\begin{eqnarray}
\label{GS_MHD}
	&\ &\hat L\Psi=\mathcal{S}_{\rm EM}+\mathcal{S}_{\rm MT}\ .
\end{eqnarray}
The differential operator $\hat L$ and the two source terms $\mathcal{S}_{\rm EM},\mathcal{S}_{\rm MT}$ are defined by
\be
\label{GS_MHD_L}
\begin{aligned}
	\hat L\Psi&= \left[\Psi_{,rr} + \frac{\sin^2\theta}{\Delta} \Psi_{,\mu\mu} \right]\ \mathcal{A}(r,\theta;\Omega) \\
	 &+\left[ \Psi_{,r} \partial^\Omega_r  + \frac{\sin^2\theta}{\Delta} \Psi_{,\mu} \partial^\Omega_\mu \right] \ \mathcal{A}(r,\theta;\Omega)\\
	& + \frac{1}{2} \left[ (\Psi_{,r})^2 + \frac{\sin^2\theta}{\Delta} (\Psi_{,\mu})^2 \right] D_\Psi^\perp\Omega\ \partial_\Omega \mathcal{A}(r,\theta;\Omega)\\
	& - \left[ (\Psi_{,r})^2 + \frac{\sin^2\theta}{\Delta} (\Psi_{,\mu})^2 \right] \frac{D^\perp_\Psi\eta_m}{\eta_m}\ \mathcal{M}^2(r,\theta)\ ,\\
	\mathcal{S}_{\rm EM} &=\frac{\Sigma}{\Delta} I D^\perp_\Psi I\ , \\
     \mathcal{S}_{\rm MT} &=-4\pi\Sigma\sin^2\theta \rho (u^tD_\Psi^\perp u_t + u^\phi D_\Psi^\perp u_\phi)\ ,
\end{aligned}
\ee
where $\mu=\cos\theta$, $I=4\pi(\mL-\eta_m u_\phi)$,
and $\mathcal{A}(r,\theta;\Omega)=-k_0(r,\theta;\Omega)+\mathcal{M}^2(r,\theta)$.
We define $\partial^\Omega_A(A=r,\mu)$ as the partial derivative with respect to coordinate $A$ with $\Omega$ fixed, $\partial_\Omega$ as the derivative with respect to $\Omega$, $D_\Psi^\perp$ as the derivative perpendicular to field lines
\footnote{$D_\Psi^\perp$ is the ordinary derivative ${\rm d}/{\rm d}\Psi$ when acting on functions of $\Psi$.}
\begin{eqnarray}
	D_\Psi^\perp&\equiv& \frac{F^A_{\ \phi}\partial_A}{F_{C\phi}F^C_{\ \phi}}\ .
\end{eqnarray}

\subsection{Characteristic Surfaces and Plasma Loading Zone}
As already shown in Paper I, the global solution of the jet structure is largely determined by
several characteristic surfaces which are related to
the singular points of the magnetic field lines and the flow velocity
\citep[see details in e.g., ][]{Beskin09,Camen86a,Camen86b,Michel69,Michel82,Takahashi90}.
The inner and outer \Alfven surfaces are the singular surfaces of the MHD GS equation (\ref{GS_MHD}),
\be
    - k_0 + \mathcal{M}^2 \big|_{r=r_{\rm A}} = 0 \ .
\ee
In the FFE limit, the \Alfven surfaces degrade to the inner and outer light surfaces (LSs),
\be
k_0  \big|_{r=r_{\rm LS}} = 0 \ .
\ee
The inner and outer fast magnetosonic (FM) surfaces are locations where the fluid speed matches the FM speed and  mathematically are determined by where the denominator of $D_\Psi^\parallel u_p$ vanishes.
The stagnation surface is where the gravity force is balanced against the centrifugal force, and mathematically is determined by
\be
D_\Psi^\parallel k_0 \big|_{r=r_*} = 0\ .
\ee
Last but not least, the null surface where the charge density is zero,
\be
{F^{t\nu}}_{;\nu}\big |_{r=r_0} =0\ ,
\ee
is not a singular surface, but is closely related to plasma production within the jet.

Recent particle-in-cell (PIC) simulations \citep{CY19} show that the electron/positron production via $\gamma-\gamma$ collision is always ignited from the null surface $r_0$.  In the MHD framework, there is no way for us to consistently take account of the detailed pair production processes, therefore we adopt a phenomenological description of the plasma loading as follows. In our MHD jet model, we plant a plasma loading zone which separates the inflow and the outflow (see Fig.~\ref{fig:cartoon}), and prescribe the plasma loading rate for the inflow/outflow by,
\be\label{sigM}
	4\pi |\eta_m|_{\rm in;out} =\frac{B_{p,0}}{\sigma_0^{\rm in;out}}\ ,
\ee
where $\sigma_0^{\rm in;out}$ is a pair of dimensionless magnetization parameters
and $B_{p,0}$ is the poloidal field on the null surface.
The rest mass flux per magnetic flux $\eta_m(r,\theta)$ is completely determined by parameters
$\sigma_0^{\rm in;out}$. Note that $\eta_{m,\rm in}<0$ and $\eta_{m,\rm out}>0$.

Thus, the energy flux per magnetic flux $\mE$ is related to the initial velocity
$u_{p,0}$ with which the plasma is injected by
\be \label{eq:Estag}
\left(\frac{\mE}{\eta_m} \right)^2 = \frac{u_{p,0}+1}{U_g(u_{p,0};\ r,\theta)}\Bigg|_{r=r_0}\ .
\ee
There is some ambiguity in $u_{p,0}$ which is the bulk velocity in the fluid description, while the fluid
description is not sufficient in the loading zone, where particles are produced without a unified moving direction.
Fortunately, we find that the velocity outside this region, especially the outflow, is insensitive to the initial velocity (see Appendix~\ref{sec:null}). Therefore, it is convenient and valid to set $u_p\equiv0$ throughout the loading zone.



\section{Numerical Results} \label{sec:results}

The global solution of a steady and axisymmetric MHD jet powered by a spinning BH
can be uniquely determined with prescribed plasma loading into the jet
and the poloidal shape of the outmost magnetic field line.

In this paper, the plasma loading is specified by the magnetization parameter $\sigma_0$ defined in Equation~(\ref{sigM}),
with $\sigma_0^{\rm in}=\sigma_0^{\rm out}$ assumed for simplicity.
The poloidal shape of outmost magnetic field line is specified by
\be\label{eq:wall}
	1=\frac{r+r_0}{r_{\rm H}+r_0}(1-\cos\theta) + \frac{1}{4} \varepsilon_{\rm vt} r\sin\theta\ .\\
\ee
This is a collimated parabola, with the 1st term being a pure parabola,
and the 2nd term being a with perturbative component \citep{Beskin98}.
In this paper, we fix the jet width parameter $r_0\equiv10 M$ and treat $\varepsilon_{\rm vt}$ as a free parameter of the jet shape.
Therefore, the free parameters in our MHD jet model are the BH spin $a$, the perturbation parameter of jet shape $\varepsilon_{\rm vt}$,
and the magnetization parameter $\sigma_0$.

Solving the MHD GS equation (\ref{GS_MHD}) is  an eigenvalue problem, with eigenvalues $\Omega(\Psi)$
and $\mL(\Psi)$ to be determined ensuring field lines cross
the inner/outer \Alfven surface smoothly.
The Bernoulli equation (\ref{eq:Bern}) is also an eigenvalue problem,
with eigenvalues $(\mE/\mL)_{\rm in}(\Psi)$ and $(\mE/\mL)_{\rm out}(\Psi)$ determined ensuring the inflow /outflow smoothly cross the inner/outer FM surface.
We numerically solve the two coupled equations iteratively.
In each iteration, the GS equation provides the field line configuration fed into the Bernoulli equation.
The stream velocity $u^\mu$ obtained from the Bernoulli equation,
are used to calculate  $\mathcal{S}_{\rm EM}$ and $\mathcal{S}_{\rm MT}$,
which are fed into the GS equation (\ref{GS_MHD}) for the next iteration.
With all quantities converge to an expected precision,
both the EM field and four-velocity $u^\mu$
of an MHD jet are obtained (see details in Appendix~\ref{sec:tech}).

\subsection{A Fiducial Model} \label{sec:fid}

In this subsection, we analyze a fiducial model with a BH spin $a=0.998$,
the perturbation parameter of the outmost field line shape $\varepsilon_{\rm vt}=0.01$,
and the magnetization parameter $\sigma_0=600$ as an example,
and summarize some generic features of MHD jets powered by spinning BHs.

\begin{figure}[h!]
\centering
\hspace{-15mm}
\includegraphics[scale=0.65]{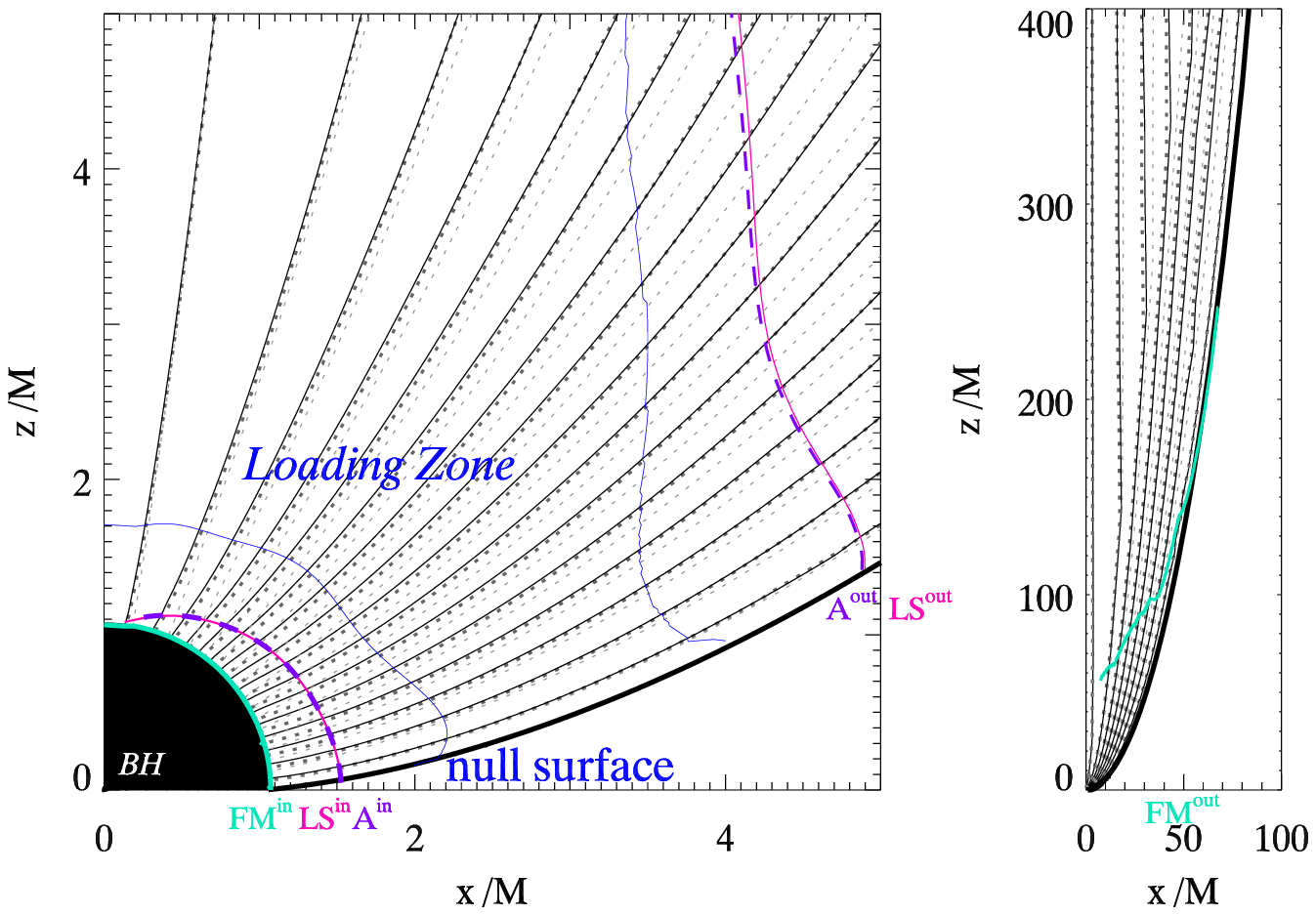}
\caption{ {\it Left:} The poloidal field line configuration of an MHD jet of
the fiducial model (thin solid lines), its FFE counterpart (thick dotted lines),
and the initial guess (thin dotted lines) for comparison.
The corresponding parameters are $(a,\varepsilon_{\rm vt},\sigma_0)=(0.998,0.01,600)$.
The outmost field line is shown in thick solid line.
The loading zone locates between the two blue lines, with the left boundary at the null surface.
The dashed purple, magenta, and aqua line represent
the Alfv\'en surfaces (A), light surfaces (LS),
and inner fast magnetosonic (FM), respectively.\\
{\it Right:} A zoom-out configuration of field lines showns in the left panel.
The outer fast magnetosonic (FM) surface is presented in aqua line.
\label{fig:FL}}
\end{figure}

\subsubsection{Magnetic Field Lines}
In Figure~\ref{fig:FL}, we compare the poloidal magnetic field lines of
the initial guess (thin dotted lines), the FFE solution (thick dotted lines), and the MHD solution (thin solid lines).  The thick lines in both panels represent the outermost field line, which is given by Equation~(\ref{eq:wall}). As expected, the deviation of the MHD solution from the FFE counterpart is relatively small, since the value of $\sigma_0$ is large (the larger $\sigma_0$, the more magnetically dominated).

In the zoom-in plot (left panel), the locations of the inner FM surface (aqua),
\Alfven surfaces (purple), and light surfaces (magenta) are presented.
For the fiducial model with a BH spin $a = 0.998$,
the outer light surface radius of the outermost field line is $\sim 5.1 M$. This radius
will increase if the BH rotates slowly, e.g., it is $\sim 9.1M$
for $a = 0.85$ as discussed later.
The loading zone locates between the two blue lines, with the left boundary set by the null surface
and the right boundary set by where $u_p|_{\rm outflow} =0$ obtained from solving the Bernoulli equation.
In the zoom-out plot (right panel), the outer FM surface terminates on some field line close to the polar direction.
Beyond this field line there exists no global solution of super-magnetosonic outflow (see details in Appendix~\ref{sec:collimj}). Fortunately, we find the fraction of such field lines is small and we set the outflow velocity as that along the last field line carrying super-magnetosonic outflow.

\begin{figure}[h!]
\centering
\includegraphics[scale=0.5]{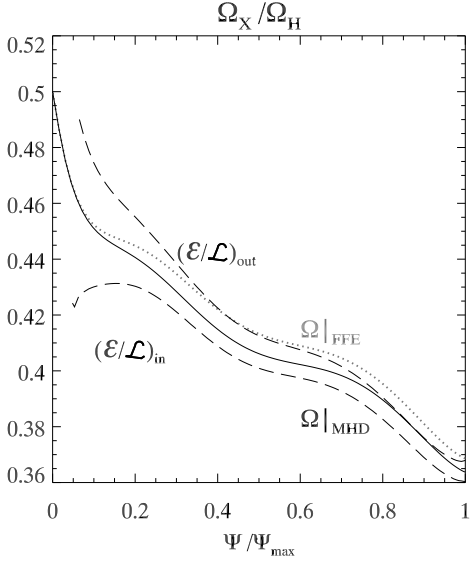}
\caption{ Comparison of angular-velocity-like quantities $\Omega|_{\rm FFE}$ (dotted), $\Omega|_{\rm MHD}$ (solid),
and $(\mE/\mL)_{\rm in;out}$ (dashed), of the fiducial model.
\label{fig:Om}}
\end{figure}

In Figure~\ref{fig:Om}, the angular velocity of magnetic field lines $\Omega|_{\rm MHD}$ (solid line),
the corresponding force-free solution $\Omega|_{\rm FFE}$ (dotted line),
and the eigenvalues $(\mE/\mL)_{\rm in;out}$ (dashed lines) of the fiducial model, are presented.
Compared to the monopole solution, in which $\Omega|_{\rm FFE}\approx0.5\Omega_{\rm H}$ (see Paper I),
the magnetic field lines of the collimated jet rotate slower.
Specifically, the outmost field line rotates with angular velocity $\Omega|_{\rm FFE}\approx0.37\Omega_{\rm H}$, consistent with previous work \citep{Nathan14,Mahlmann18}.
In addition, the fluid inertia drags down the rotation of all magnetic field lines further, i.e., $\Omega|_{\rm MHD}<\Omega|_{\rm FFE}$.

\subsubsection{Flow Pattern}

\begin{figure}[h!]
\centering
\includegraphics[scale=0.9]{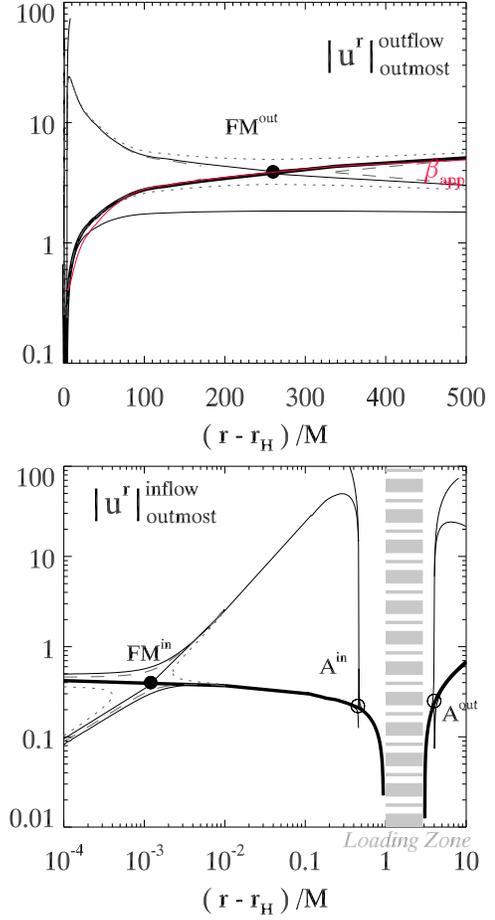}
\caption{ {\it Top:} The fluid velocity $|u^r|$ of the outflow carried on the outmost field line of the fiducial model.
The outer FM point is marked by filled circle.
The solid lines show the solutions with correct eigenvalues $(\mE/\mL)_{out}$.
The dashed and dotted lines show the solutions with eigenvalues of slightly off the correct values.
The corresponding superluminal apparent speed $\beta_{\rm app}$ is presented in red line in the top panel.
A jet viewing angle of $i=17^\circ$ is adopted.\\
{\it Bottom:} The fluid velocity $|u^r|$ of inflow.
The solutions with correct eigenvalues $(\mE/\mL)_{\rm in}$ are shown in solid lines
and the ones with eigenvalues of slightly off the correct values are shown in dashed and dotted lines.
The inner FM point and the Alfv\'en points are respectively marked by filled circle and open circles.
The gray shaded region represents the loading zone.
\label{fig:ur}}
\end{figure}

In Figure~\ref{fig:ur}, we show the outflow/inflow velocity $u^r$ along the outermost field line
in top/bottom panel, where the outflow velocity $u^r$ is approximately equal to the Lorentz factor $\gamma$
at large distance from the central BH.

We show both the physical numerical solutions and and the unphysical ones to the Bernoulli equation (\ref{eq:Bern}), where the physical ones are emphasized in thick lines.
To clarify the `X'-type of the FM point, we show the solutions with $(\mE/\mL)_{\rm in/out}$
slightly off the correct values in dashed and dotted lines, respectively.
For the inflow, there exist multiple roots only for
correct eigenvalue $(\mE/\mL)_{\rm in}$, while there is no global solution, with slightly low $(\mE/\mL)_{\rm in}$ (dotted line in bottom panel) and there exist no multiple roots, with slightly high $(\mE/\mL)_{\rm in}$ (dashed line in bottom panel). The similar is also true for the outflow.

In the top panel, we also present in red line the superluminal apparent speed $\beta_{\rm app}$, calculated by:
\be\label{eq:appv}
	\beta_{\rm app} = \frac{\beta_V\sin i}{1-\beta_V\cos i} \ ,
\ee
where $\beta_V= \sqrt{1-\gamma^{-2}}$ with the Lorentz factor $\gamma = -u_t$ for the outflow.
As an application to the M87 jet, the jet viewing angle is
chosen to be  $i=17^\circ$  \citep{Mertens16a}.

\begin{figure}[h!]
\centering
\includegraphics[scale=1]{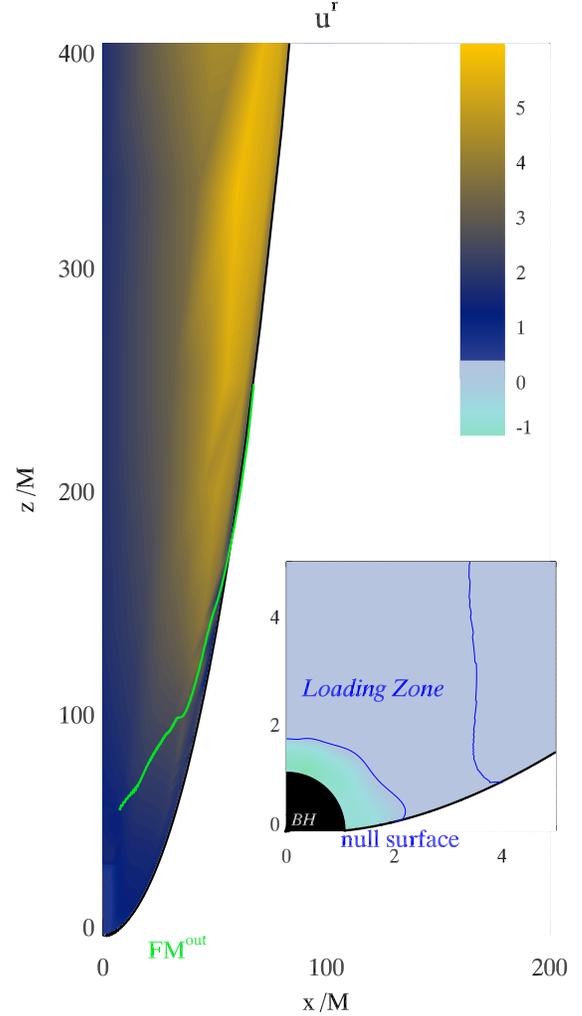}
\caption{ Corresponding 2-dimensional contour of the fluid velocity field $u^r$ of the fiducial model,
with a zoom-in version which illustrates the inflow configuration and the loading zone in the sub-panel.
The contra-variant radial component of velocity is approximately equal to the Lorentz factor, i.e., $u^r\approx\gamma$,
at large distance from the central BH.
\label{fig:ur2d}}
\end{figure}

In Figure~\ref{fig:ur2d}, we show contour of the fluid velocity field
$u^r$ which displays clear stratification with
fluid moving relatively faster in the `sheath' (close to the outmost field line) and slower in the `spine' (close to the polar axis). Shown in the zoom-in contour plot is the inflow pattern which is more isotropic.

\subsubsection{Comparison with Previous Studies}
\begin{figure}[h!]
\centering
\includegraphics[scale=0.5]{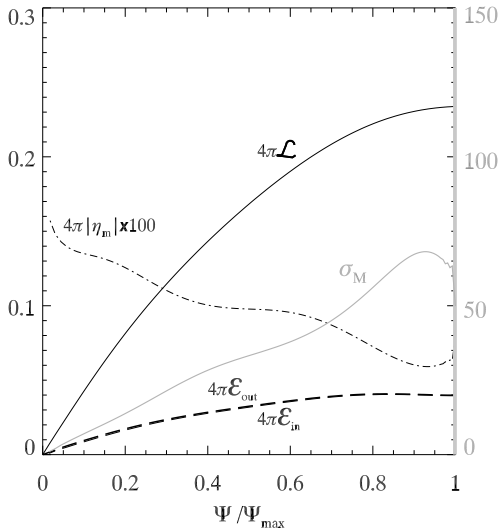}
\caption{The energy flux in outflow and inflow $4\pi\mE_{\rm out;in}$ (thick and thin dashed black lines), the angular momentun flux $4\pi\mL$ (solid black line),
and the rest mass flux $4\pi|\eta_m|$ (dashed-dotted black line) of the fiducial model are shown with respect to the left y-axis.
The Michel's magnetization parameter $\sigma_{\rm M}$ (solid gray line) is shown with respect to the right y-axis.
\label{fig:edot}}
\end{figure}

In some previous studies, the fluid acceleration within a magnetically dominated jet has been investigated in various context \citep[e.g.,][]{Li1992, Beskin98, Vlahakis2003, Vlahakis2004, Beskin06,Tchek08,Komi09}.  To compare with them,  we first define the Michel's magnetization parameter \citep{Michel69} $\sigma_{\rm M}$ as
\be
\sigma_{\rm M}(\Psi)=\frac{\mE_{\rm out}(\Psi)}{\eta_m(\Psi)}\ .
\ee
This parameter quantifies the maximal Lorentz factors acquired by the fluid if all the Poynting flux was converted into the fluid kinetic energy. According to the previous studies, the fluid is linearly accelerated in
the subsonic regime, i.e., $\gamma\propto \varpi :=r\sin\theta$, until the FM point $ \varpi_{\rm FM} \sim \sigma_{\rm M}^{1/3}/\Omega$ and $\gamma_{\rm FM}\simeq \sigma_{\rm M}^{1/3}$, while the acceleration in the supersonic regime is ascribed to the action of ``magnetic nozzle". As detailed by \cite{Vlahakis2003, Vlahakis2004},
the acceleration is only possible if $\partial (B_{\rm T} \varpi)/\partial l <0$, where $B_{\rm T}$ is the toroidal magnetic field component and $l$ is the coordinate along the magnetic surface.

In Figure~\ref{fig:edot}, we show the energy flux $4\pi\mathcal{E}$, the momentum flux $4\pi\mathcal{L}$,
the fluid rest mass flux $4\pi\eta_m$ and the Michel's magnetization parameter $\sigma_{\rm M}$ as functions of  $\Psi$.
Specifically,   $\sigma_{\rm M}\approx 70$ on the jet sheath and the corresponding Lorentz factors at the outer FM points $\gamma_{\rm FM}\approx4$ (Figure~\ref{fig:ur}), which is consistent with the analytical estimate $\sigma_{\rm M}^{1/3}$.
As for the location of outer FM surface $\varpi_{\rm FM}$, our numerical result (Figure~\ref{fig:ur2d}) is consistent with the analytical estimate $\sigma_{\rm M}^{1/3}/\Omega$ within a factor of $\sim 2$. In the supersonic regime, we find
$\partial(B_{\rm T} \varpi)/\partial l \approx 0$ and the acceleration efficiency is low, i.e., the magnetic nozzle  does not operate efficiently. \footnote{To demonstrate the acceleration efficiency dependence on the field line shape, we also solve the Bernoulli equation for a field line with a much larger vertical component $\epsilon_{\rm vt} = 0.065$ and we find a linear acceleration $\gamma\propto \varpi$ far beyond the FM surface. Thereby we should not take the linear acceleration for granted and a self-consistent treatment of both the field lines and the fluid acceleration is necessary. }

In Table \ref{tab:1}, we summarize the  of the jet properties of the fiducial model, where
we use $\dot{E}_{\rm MT}(r_{\rm max})$ to estimate $\dot{E}_{\rm MT}^\infty$ considering the low acceleration efficiency in the supersonic regime.
For the fiducial model, the matter component takes away $\approx20\%$ of total energy flux at infinity.
Only $\approx1\%$ of total energy flux would be taken away by matter,
with the same $a$ and $\sigma_0$, but in monopole field lines carrying sub-magnetosonic outflows (Paper I).
Hence, the energy conversion from EM field to matter is much more efficient in the super-magnetosonic outflow.

\subsection{Application to the M87 jet}

In this subsection, we apply our MHD model to the M87 jet aiming to explain the
apparent speed measurements in the range of $r\in [10^2, 10^5] M$.

\subsubsection{Observational Facts}

\begin{figure}[h!]
\centering
\hspace{-15mm}
\includegraphics[scale=0.65]{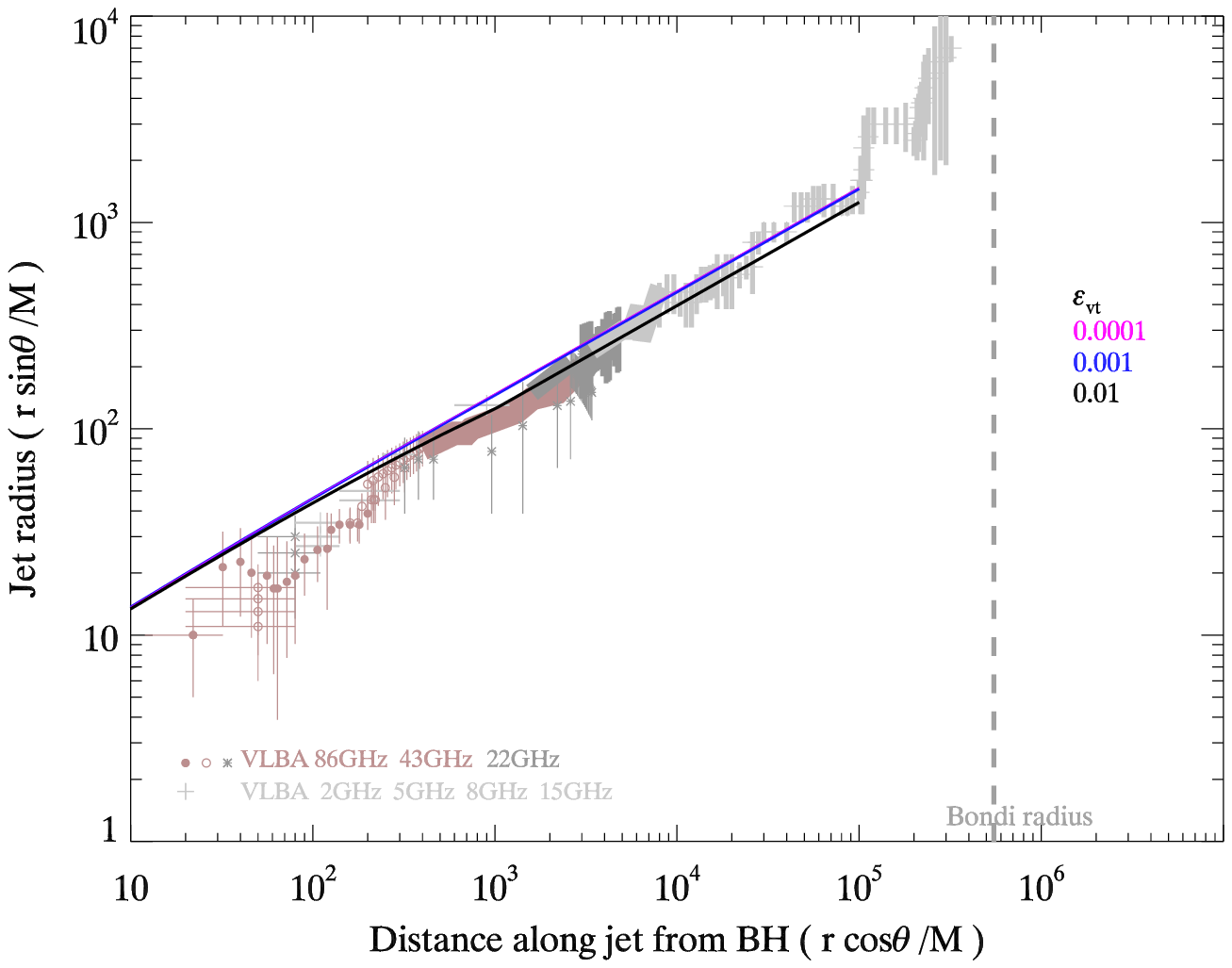}\\
\hspace{-15mm}
\includegraphics[scale=0.65]{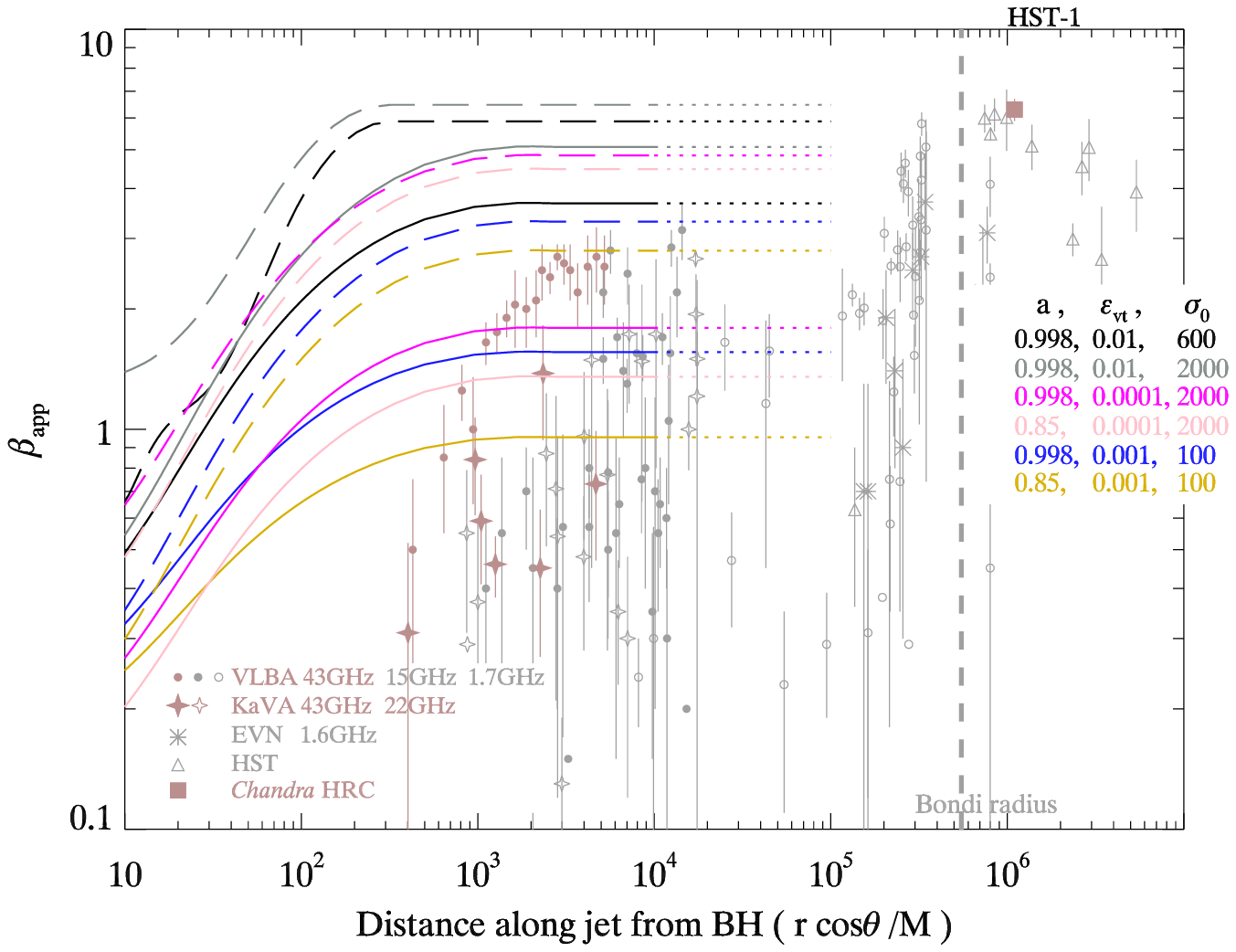}
\caption{{\it Top}:  jet radius ($\varpi=r\sin\theta$) profile as a function of axial distance ($z=r\cos\theta$).
The black, blue, and magenta lines represent profiles of the outmost field lines of
pure paraboloidal magnetospheres with $\varepsilon_{\rm vt}=0.01$, $0.001$, and $0.0001$, respectively.
The location of Bondi radius is shown in thick dashed line.
Observational data are selected from \citet{Hada11,Hada13,Hada16}. \\
{\it Bottom}: Apparent superluminal speed $\beta_{\rm app}$ from the fiducial model (black)
and five models with different parameters (gray, magenta, pink, blue, and yellow).
In each model, the upper and lower boundaries of  $\beta_{\rm app}$ are shown in dashed and solid line, respectively.
The filled and open circles represent VLBA data \citep{Cheung07,Mertens16,Park19}.
The filled and open diamonds represent KaVA data  \citep{Park19},
The stars, triangles and filled square represent data from EVN, HST, and {\it Chandra} HRC, respectively \citep{Asad14,Bire99,Snio19}.
\label{fig:jetw}}
\end{figure}

The structure of M87 jet has been found to be in an edge-brightened parabolic profile
in the range of $\sim10^2-10^5M$ from the central BH.
The latest VLBA observations \citep{Hada11,Hada13,Hada16} provide a best fit for the jet width as $W_{\rm jet}\propto r^{0.56\pm0.03}$ (see the upper panel of Figure~\ref{fig:jetw}), which we take as an input to our MHD jet model.
Overlaid on the data points are three jet profiles defined in Equation~(\ref{eq:wall}) with $\varepsilon_{\rm vt}=0.01$ (fiducial model), $0.001$, and $0.0001$, respectively.

Latest observations of HST-1 show clear evidence of superluminal motion with apparent speed of $\sim6.3c$ \citep{Snio19}.
However, the speed of the inner jet upstream of HST-1 was found to be slow \citep[][etc.]{Bire99,Cheung07,Asad14}
and stratified \citep{Mertens16,Park19}, i.e., at each radius in the range of $\sim10^2-10^5M$,
the apparent speed extends a wide range, from $<0.1c$ to $3c$ (see the bottom panel of Figure~\ref{fig:jetw}).

\subsubsection{Previous Interpretation}

The observed velocity stratification has been interpreted as the result of either a two-fluid jet,
or a jet within which an instability mode activated. In the former explanation,
a fast, accelerating sheath originates from the BH magnetosphere or from the inner part of the accretion disk,
and a non-relativistic wind originates from the outer part of the disk \citep{Sol89,Tsin02,Komi07,Yuan15}.
In the latter, a slow pattern associated with the current driven instability or the Kelvin-Helmholtz instability
develops on the fast jet sheath \citep{Hardee2000, Lobanov2003}.
The subluminal component can be explained by either model, while the superluminal component is still not well understood,
because it is believed to reflect the true speed of the jet, while the flow acceleration along the jet is not clear yet.

Particle drift velocity in the jet EM field proposed by \cite{NMF07} has been widely used to interpret the jet kinematics
\citep[e.g.,][]{Penn13,Naka18,Lucc19},
i.e., $v=|{\mathbf E}\times {\mathbf B}|/B^2$. With the FFE assumption, \cite{NMF07}
found a simple linear relation $\gamma(r)\propto \varpi(r)$, which is consistent with e.g., \citet{Beskin06} for jets within which
the magnetic nozzle operates efficiently in the supersonic regime.
In a recent study, \cite{Naka18} calibrated the power law $\gamma \propto \varpi(r)\propto r^{0.56}$
for the M87 jet with GRMHD simulations and found it
greatly overestimates the apparent speed profile than what has been observed. A likely reason that this simple scaling fails is the low acceleration efficiency in the supersonic regime of the M87 jet as discussed in the previous section. 
In summary, there is still no viable explanation to the jet kinematics of M87, especially to the superluminal part.

\subsubsection{Interpreting the Apparent Speed Profile with Our MHD Model}

As shown in Figure~\ref{fig:ur} and \ref{fig:ur2d}, the outflow is rapidly accelerated inside the outer FM surface,
and the acceleration significantly slows down outside the outer FM surface.
We calculate the apparent speed along each field line $\beta_{\rm app}$ using Equation~(\ref{eq:appv}),
and show the fastest (dashed) and the slowest (solid) in the bottom panel of Figure~\ref{fig:jetw}.
In practice, we find that the fastest line is located in the jet sheath, while the slowest one is in the jet spine.
We horizontally extend the apparent speed from $10^4M$, the outer boundary of our computational domain, to $10^5M$
since no significant acceleration is expected in this region.\footnote{Outside $10^5 M$, the radial profile of the jet
width displays an abrupt change, which is an indicator of some different physical process kicking in.}

\begin{table*}
\centering
\begin{tabular}{|c|ccc|cccccc|}
 \hline
       & $a$ & $\varepsilon_{\rm vt}$ & $\sigma_0$ &  $\dot{E}_{\rm Poynting}^\infty$ & $\dot{E}_{\rm MT}^\infty$ & $\dot{E}_{\rm MT}^\infty/\dot{E}_{\rm tot}^\infty$ & $\dot{M}_{\rm out} $ & $\dot{M}_{\rm out} /\dot{E}_{\rm tot}^\infty$ & $\beta_{\rm app}|_{r=10^4 M}$  \\
 \hline
 Fiducial Model & $0.998$  & $0.01$  & $600$ & $0.0233$ & $0.0059$ & $20\%$ & $0.00097$ & $3.3\%$ & $(3.7, 5.9)$  \\
             Case II  & $0.998$  & $0.01$ & $2000$ & $0.0267$ & $0.0040$ & $12\%$ & $0.00032$ & $1\%$ & $(5.1, 6.5)$ \\
             Case III & $0.998$  & $0.0001$ & $2000$ & $0.0255$ & $0.0030$ & $10\%$ & $0.00028$ & $1\%$ &$(1.8, 4.8)$ \\
             Case IV  & $0.85$   & $0.0001$ & $2000$ & $0.0112$ & $0.0015$ & $12\%$ & $0.00019$ & $1.5\%$ &$(1.4, 4.5)$ \\
             Case V&  $0.998$  & $0.001$ & $100$ & $0.0173$ & $0.0115$  & $39\%$ & $0.0057$ & $20\%$ &$(1.6, 3.3)$ \\
             Case VI &  $0.85$   & $0.001$ & $100$ & $0.0075$ & $0.0057$  & $43\%$  & $0.0039$ & $30\%$ &$(0.95, 2.8)$ \\
 \hline
\end{tabular}
\caption{\label{tab:1} A summary of the examples considered in this paper, where
the EM $\dot{E}_{\rm Poynting}^\infty$ and matter $\dot{E}_{\rm MT}^\infty$ components of energy extraction rate at infinity,  and the rest mass outflow rate $\dot{M}_{\rm out} $ are shown in units of $\Psi_{\rm max}^2$.  The total energy extraction rate $\dot{E}_{\rm tot}^\infty=\dot{E}_{\rm Poynting}^\infty+\dot{E}_{\rm MT}^\infty$ (see Paper I for explicit definitions of $\dot E_{\rm tot, Poynting, MT}$),
and the plasma loading rate $\dot M_{\rm load} = \dot{M}_{\rm out} + \dot{M}_{\rm in}$ with $\dot M_{\rm in; out} := \int_0^{\Psi_{\rm max}} 4\pi|\eta_m|_{\rm in; out}\ {\rm d}\Psi $.
}
\end{table*}

Though the fiducial model successfully reproduces both the slow acceleration (outside the FM surface)
and the jet velocity stratification, the predicted radial profile of the apparent speed
is systematically higher than what has been observed. To understand why,
we need to figure out how the BH spin ($a$), the plasma loading rate ($\sigma_0$) , the shape of the outmost field line ($\epsilon_{\rm vt}$) affect the fluid acceleration. Intuitively, it is straightforward to expect more collimated and magnetically dominated jets
powered by more rapidly spinning BHs to drive faster outflow, and here we are to quantify the dependence via numerical experiments. For this purpose, we explore more parameter space and collect 5 typical
cases in Table~\ref{tab:1} and Figure~\ref{fig:jetw}.

The role of BH spin is transparent: the faster the BH spins, the higher energy extraction rate in the form
of Poynting flux and the higher mode speeds, especially the FM speed, consequently the higher flow speed (see Case III versus IV, and V versus VI). Plasma loading rate is pulling in the opposite direction: the more matter loaded, the slower the jet rotates, the lower FM speed and the lower energy available per unit matter, consequently the lower flow speed (see Case II versus the fiducial model). The outflow acceleration efficiency is also sensitive to the jet shape.
The apparent speed profile of Case III is much lower than that of II, which is slightly different from the former case in the jet shape with smaller vertical component. This sensitive dependence on the jet shape is unexpected in the particle drift velocity interpretation.

From above analysis, the slow outflow is possibly a result of a slowly spinning BH, a heavily loaded jet or a jet
shape of low acceleration efficiency (small $\epsilon_{\rm vt}$ in our model).
For a BH with a nearly extremal spin (Case II and III), it seems unlikely to explain the slow outflow with the jet shape alone because the outflow apparent speeds become convergent when we examine jet shapes with even smaller vertical component.
The slow outflow can be explained by a heavily loaded jet as in Case V and VI, where the rest mass outflow rate is $20\%-30\%$
of the total energy extraction rate. Currently, the most promising plasma loading mechanism is the electron/positron pair production from $\gamma-\gamma$ collisions at the jet base, while latest PIC simulations show the power dissipated in the pair production gap is $\lesssim (0.1\%-1\%)$ for the M87 jet \citep{CY19}. Therefore the role of plasma loading in the slow outflow is also limited.
To conclude this section, the slow outflow might be a result of a small BH spin, which has not been well constrained yet.

\section{Summary and Discussion}\label{sec:summary}

\subsection{Summary}

We constructed a full MHD model of BH jets,
in which EM fields and fluid motions are respectively governed
by the MHD GS equation (\ref{GS_MHD}) and the Bernoulli equation (\ref{eq:Bern}).
From the two governing equations, the Maxwell tensor $F_{\mu\nu}$, the fluid rest mass density $\rho$, and the fluid four-velocity $u^\mu$,
can be completely determined with a proper poloidal shape of the outmost field line
and a magnetization parameter $\sigma_0$ given.
To be consistent with recent PIC simulations \citep{CY19} which show that the electron/positron pair production via $\gamma-\gamma$ collisions in a gap around the null surface is an effective mechanism of loading plasma into the BH jet, we assume a plasma loading zone with the inner boundary on the null surface and the outer boundary self-consistently determined by the condition $u_p=0$, i.e.,
we take the starting point as the outer boundary of the loading zone. In summary, there are 3 parameters in our MHD jet model: the dimensionless BH spin $a$, the magnetization parameter $\sigma_0$ and $\epsilon_{\rm vt}$ controlling the jet shape.

Different from Paper I, we focused on collimated jets and the jet kinematics in this paper.
The Bernoulli equation is solved for the inflow and outflow as eigenvalue problems, respectively, where
the eigenvalues $(\mE/\mL)_{\rm in}$ and $(\mE/\mL)_{\rm out}$ are determined ensuring the inflow/outflow smoothly cross
the inner/outer FM surface.
We found that the outflow is always sub-magnetosonic if the poloidal shape of outmost field line is a pure parabole (see Figure~\ref{fig:collim}),
and the super-magnetosonic solution exist only if the jet is more collimated than the pure parabole which we parameterize with a vertical perturbative component (Equation~\ref{eq:wall}).
The MHD GS equation is also a problem to be solved
by searching correct eigenvalues for $\Omega(\Psi)$ and $\mL(\Psi)$
to ensure field lines smoothly cross the inner/outer \Alfven surface.
We numerically solved the two equations and acquire both the EM fields $F_{\mu\nu}$ and the flow velocity $u^\mu$ with all eigenvalues obtained (see details in Appendix~\ref{sec:tech}).

As an example, we consider a fiducial model with parameters $(a,\varepsilon_{\rm vt},\sigma_0)=(0.998,0.01,600)$,  from which we find the rotation of field lines is obviously dragged down by the plasma loaded, i.e., $\Omega|_{\rm MHD} < \Omega|_{\rm FFE}$.
The fluid moves relatively faster in the jet sheath (near the outmost field line)
but slower in the jet spine (near the polar axis).
The loss of the Poynting flux is converted into the fluid kinetic energy.
The matter component takes away $\approx20\%$ of the Poynting flux at infinity.
The energy conversion from EM field to fluid motion is much more efficient in the super-magnetosonic outflow,
than in the sub-magnetosonic one carried by field lines in monopole configuration (Paper I).

In the fiducial model, both the outflow velocity stratification and slow acceleration outside the FM surface are consistent with
what has been observed in the M87 jet. But the fiducial model systematically overestimates the outflow apparent speeds.
The slow outflow is possibly a result of a slowly spinning BH, a heavily loaded jet or a jet
shape of low acceleration efficiency (small $\epsilon_{\rm vt}$ in our model). With some numerical experiments, we found
the jet shape has limited effect in reducing the outflow acceleration, and outflow as slow as observed in the M87 jet is possible only if the plasma loading is orders of magnitude higher than the expected level from pair production at the jet base, therefore a slowly spinning BH is likely responsible for the slow outflow.

\subsection{Discussion}

In this paper, we assume a plasma loading zone consistent with the mechanism of pair production via $\gamma-\gamma$ collisions
in a gap at the jet base \citep{Lev11, Brod15, Hiro16, Chen18, CY19}. Another possible source of the high-energy photons is the
inner part of the accretion disk, which, as shown by \cite{Lev11}, is not enough to fill the M87 jet with plasma of Goldreich-Julian density.

Is it possible that a considerable part of matter within the jet comes from the accretion disk or wind?
The Larmor radii of electrons and protons in M87 jet loading zone
are estimated to be $r_{{\rm L};e}=\gamma_e m_e v_e/(e|B|)\approx2.8\times10^{-3}$km and
$r_{{\rm L};i}=\gamma_i m_i v_i/(e|B|)\approx0.8$km, respectively.
$v_{e,i}$ and $\gamma_{e,i}$ are the thermal velocity
of electrons/protons and the corresponding Lorenz factor, respectively,
according to typical temperature $T_e\sim10^{10}$ K and $T_i\sim10^{12}$ K in the inner part of the accretion disk \citep{Yuan14}.
The averaged magnetic field strength is chosen to be $|B|=20$G, estimated from recent simulations \citep[e.g.,][]{EHT19V,Chael19}.
The Larmor radii are extremely small compared with the gravitational radius of M87 BH ($\approx10^{10}$km).
Therefore, it is hard for particles from the disk to be injected into the jet.
A recent large-scale GRMHD simulation work \citep{Chatter2019} implied that the interaction between the jet and the disk wind leads to the development of pinch instabilities, via which matter is loaded into the jet.
In this scenario, the slow outflow of the M87 jet is a result of a large amount of matter loaded and heated by the dissipation of the magnetic field. The development of jet instabilities is not yet well understood on the theory side \citep{Kim2018}, while it predicts a quite different radial profile of the matter density $\rho(r)$ along the jet, which we plan to constrain from previous observations of the M87 jet.

\section*{Acknowledgements}
We thank the referee for his/her valuable suggestions and comments. We also thank Hung-Yi Pu for very helpful discussions.
L.H. thanks the support by the Key Program of the National Natural Science Foundation of China (grant No. 11933007), the National Natural Science Foundation of China (NSFC grants 11590784 and 11773054), Research Program of Fundamental and Frontier Sciences, CAS (grant No. ZDBS-LY-SLH011), and Key Research Program of Frontier Sciences, CAS (grant No. QYZDJ-SSW-SLH057).
Z.P. is  supported by Perimeter Institute for Theoretical Physics.  Research at
Perimeter Institute is supported in part by the Government
of Canada through the Department of Innovation,
Science and Economic Development Canada and by the
Province of Ontario through the Ministry of Colleges and
Universities.
C.Y. is supported by NSFC (grants 11521303, 11733010 and 11873103),
and Open Projects Funding of Lunar and Planetary Science Laboratory, MUST Partner Laboratory of Key Laboratory of Lunar and Deep Space Exploration, CAS (Macau FDCT grant No. 119/2017/A3).
This work made extensive use of the NASA Astrophysics Data System and
of the {\tt astro-ph} preprint archive at {\tt arXiv.org}.

\appendix

\section{Numerical Techniques} \label{sec:tech}

In this appendix, we detail the procedure of solving the
MHD GS equation (\ref{GS_MHD}) and the Bernoulli equation (\ref{eq:Bern}) consistently,
for an MHD jet powered by a spinning central BH
and with collimated paraboloidal shape of outmost field line.
We solve the coupled governing equations iteratively:
\begin{eqnarray} \label{Eq_set}
\left\{\begin{array}{cc}
    \hat{L}\Psi^{(l)}=(\mathcal{S}_{\rm EM}+\mathcal{S}_{\rm MT})\{{\mL}^{(l)}, u^{(l-1)}, \eta_m^{(l-1)}\}\ ,& \\
    \mathcal{F}\{u^{(l)}; \Omega^{(l)}, ({\mE}/{\mL})^{(l)}, \Psi^{(l)}\} =0 \ ,&{\rm with}\ l=1,2,\cdots\ .
\end{array}
\right.
\end{eqnarray}
The shape of outmost field line is depicted by a collimated parabole,
which satisfies Equation~(\ref{eq:wall}). Thus, we adopt an initial guess of magnetic flux function as
\begin{equation}
\begin{aligned}\label{eq:collimj}
	\Psi^{(0)}(r,\theta) &= \Psi_{\rm max} \left[\frac{r+r_0}{r_{\rm H}+r_0}(1-\cos\theta) + \frac{1}{4}\varepsilon_{\rm vt} r\sin\theta\right]\ .\\
\end{aligned}
\end{equation}
We activate the iteration with $\Psi^{(0)}(r,\theta)$ and
\begin{eqnarray} \label{init}
\left\{\begin{array}{cccc}
    \Omega^{(0)}(\Psi)&=&0.5\Omega_{\rm H}\ ,  \\
    {\mathcal L}^{(0)}(\Psi)&=&\Omega_{\rm H} \Psi[2-(\Psi/\Psi_{\rm max})]/(8\pi)\ ,\\
    \eta_m^{(0)}(r,\theta) &=& u^{(0)}(r,\theta) = 0\ .
\end{array}
\right.
\end{eqnarray}
For simplicity, we specify the same magnetization parameters for inflow and outflow, i.e., $\sigma_0^{\rm out}=\sigma_0^{\rm in}$.

The numerical techniques basically follow those described in Paper I.
In addition, there are some special treatments for collimated paraboloidal field lines.
We detail the techniques and special treatments as follows:

\begin{itemize}
\item[\it Step 1]
The second-order differential MHD GS equation degrades to a first-order one
on critical \Alfven surfaces, where $\mathcal{A}(r,\theta)=0$.
For numerical convenience, a new radial coordinate $R = r/(r+1)$ is defined.
The computation domain is confined $R\times\mu=[R(r_{\rm H}), R_{\rm max}]\times[0,1]$
with a uniform $512\times128$ grid implemented.
We choose $R_{\rm max}=0.9999$, correspondingly $r_{\rm max}\approx10^4 M$, which is far enough from the BH
to search the outer FM surface.

During searching the solution of magnetosphere configuration, we evolve $\Psi$ on grids only inside the jet,
i.e., $\mu\ge\mu_{\rm outmost}(r)$, while remain the values outside unchanged.
The four boundary conditions are listed as follows:
\begin{eqnarray} \label{boundary}
\left\{\begin{array}{cccc}
    \Psi|_{\mu=\mu_{\rm outmost}(r)}&=&\Psi_{\rm max},&  \\
    \Psi|_{\mu=1}&=&0,&  \\
    \Psi_{,r}|_{r=r_{\rm H}}&=&0,& \\
    \Psi_{,r}|_{r=r_{\rm max}} &=&0;&{}_{\mu\to\Lambda=(\mu-\mu_{\rm outmost}(r))/(1-\mu_{\rm outmost}(r))}\ .
\end{array}
\right.
\end{eqnarray}
The first three are the boundary conditions on the outmost field line, in the polar direction, and at horizon, respectively.
The last one is the outer boundary, defined with the aid of an auxiliary variable
$\Lambda=(\mu-\mu_{\rm outmost}(r))/(1-\mu_{\rm outmost}(r))$.
On has $\Lambda\in[0,1]$ corresponding to the region with $\mu\in[0,\mu_{\rm outmost}(r)]$.
It should be pointed out that the outer boundary can be written in such simple form only in new coordinates $R\times\Lambda$.
Except treating the outer boundary, we stick to adopt uniform grid in $\mu$ in the differential computation of MHD GS equation.
It has advantages in two aspects. One is to avoid the cross term in the MHD GS equation.
The other is one can explore the mechanism related with disc and corona by modeling the flux function $\Psi>\Psi_{\rm max}$ in future work.

We use the overrelaxation solver with Chebyshev acceleration \citep{Press86} to evolve the flux function $\Psi^{(l)}$.
Input the initial guess $\left\{\Psi^{(l-1)},\Omega^{(l-1)},{\mL}^{(l-1)}\right\}$ from  previous loop.
$\Psi^{(l)}(r,\theta)$ is updated inside the jet ($\Psi\le\Psi_{\rm max}$),
except those in the two \Alfven surfaces vicinity.
Different flux functions on the outer \Alfven surface (OA), $\Psi(r_{\rm OA}^-)$ versus $\Psi(r_{\rm OA}^+)$, are obtained from interpolations.
We adjust $\Omega^{(l)}(\Psi)$ on OA to reduce the difference by
\begin{eqnarray}
	\Omega^{(l)}_{\rm new}(\Psi_{\rm new})&=& \Omega^{(l)}_{\rm old}(\Psi_{\rm old}) + 0.05 [\Psi(r_{\rm OA}^+)-\Psi(r_{\rm OA}^-)], \nonumber\\
	&{\rm with}&\nonumber\\
	\Psi_{\rm new}&=& 0.5[\Psi(r_{\rm OA}^+)+\Psi(r_{\rm OA}^-)]\ ,
\end{eqnarray}
where the quantities with subscript old/new are those before/after adjustment.
Similarly, we adjust both $\Omega^{(l)}(\Psi)$ and ${\mathcal L}^{(l)}(\Psi)$ at the inner Alfv\'en surface (IA):
\begin{eqnarray}
	\Omega^{(l)}_{\rm new}(\Psi_{\rm new})&=& \Omega_{\rm old}(\Psi_{\rm old}) + 0.05[\Psi(r_{\rm IA}^+)-\Psi(r_{\rm IA}^-)],\nonumber\\
	{\mathcal L}^{(l)}_{\rm new}(\Psi_{\rm new})&=& {\mathcal L}^{(l)}_{\rm old}(\Psi_{\rm old}) - 0.05[\Psi(r_{\rm IA}^+)-\Psi(r_{\rm IA}^-)],\nonumber\\
	&{\rm with}&\nonumber\\
	\Psi_{\rm new}&=& 0.5[\Psi(r_{\rm IA}^+)+\Psi(r_{\rm IA}^-)]\ .
\end{eqnarray}

The numerical techniques are similar to those
in previous work \citep[e.g.][]{Contop13,Nathan14,Mahlmann18,Pan17,Huang16,Huang18,Huang19}.
On can obtain converged $\{\Psi^{(l)},\Omega^{(l)},{\mL}^{(l)}\}$ with field lines
smoothly crossing the inner/outer \Alfven surface after sufficient evolution.

\item[\it Step 2]
With the magnetosphere configuration specified by MHD GS equation,
the Bernoulli equation~(\ref{eq:Bern}) is solved on each field line
as a problem of two eigenvalues $({\mathcal E}/{\mathcal L})_{\rm in}^{(l)}$ and $({\mathcal E}/{\mathcal L})_{\rm out}^{(l)}$.
In the Kerr space-time, $u_p$ diverges at the black hole horizon $r_{\rm H}\left(\equiv M+\sqrt{M^2-a^2}\right)$.
For convenience in the numerical calculation, we transform $u_p$ into the contravariant radial velocity $u^r$
then obtain
\begin{eqnarray}\label{eq:Bernur}
	\mathcal{F}(u)&=&(u^r)^2f_1^2+1-\left(\frac{{\mathcal E}}{\eta_m}\right)^2 U_g(r,\theta)=0\ ,
\end{eqnarray}
with a factor
\begin{eqnarray}
	f_1^2&=& g_{rr}+g_{\theta\theta}(\Psi_{,r})^2/(\Psi_{,\theta})^2 .
\end{eqnarray}
With the aid of $\sigma_0$, the Bernoulli equation (\ref{eq:Bernur}) becomes a fourth-order polynomial equation as
\begin{eqnarray} \label{eq:Bern2}
	\sum_{i=0}^4B_i (u^r)^i&=&0\ ,
\end{eqnarray}
where the coefficients $B_i$ read
\begin{equation}
\begin{aligned}
	B_4&= \frac{f_1^4}{\sigma_0^2}\frac{B_{p,0}^2}{B_p^2}\ ,\\
	B_3&= - \frac{2k_0f_1^3}{\sigma_0}\frac{B_{p,0}}{B_p}\ ,\\
	B_2&= k_0^2f_1^2 + \left(1 + \frac{{\mathcal E}^2}{\eta_m^2} k_4\right) \frac{f_1^2}{\sigma_0^2}\frac{B_{p,0}^2}{B_p^2}\ ,\\
	B_1&= \left(-k_0 + \frac{{\mathcal E}^2}{\eta_m^2} k_2\right) \frac{2f_1}{\sigma_0}\frac{B_{p,0}}{B_p}\ ,\\
	B_0&= k_0^2 - \frac{{\mathcal E}^2}{\eta_m^2} k_0k_2\ .\\
\end{aligned}
\end{equation}
With correct eigenvalues of inflow and outflow chosen on each field line,
we acquire solution $\{u^{(l)}, ({\mE}/{\mL})^{(l)}\}$
which ensures the inflow/outflow cross the inner/outer FM surface smoothly.

Compared to monopole configuration, the collimated field lines
severely deviate from straight lines $\theta\equiv\theta(r_{\rm H})$ and concentrate to the polar axis.
Here we adopt $\theta_{{\rm FL},i}$, representing values of $\theta$ related to $r_i$ with $\Psi$ conserved,
to differ from values on grids $\theta_{{\rm grid},j}$ with $\mu=\cos\theta_{{\rm grid},j}$.
Unfortunately, with uniform grid in $R\times\mu$ adopted, the resolution is relatively poor
in the regime $(r\to r_{\rm max})\times(\mu\to1)$.
In order to calculate the derivatives of the flux function precisely in the whole computational domain,
we appeal to another recipe described as follows,
instead of the central difference.

On each slice of the computation domain, where $r\equiv r_i$, we fit
\be
\begin{aligned}	
	\Psi|_{r_i}(\theta_{{\rm grid},j})&=f_i(\theta)\ ,\ j=1,2,\cdots\ ,
\end{aligned}
\ee
with a polynomial function $f_i$. The values of the poloidal derivative of flux function can be calculated by
\be\label{eq:dPsidth}
	\Psi_{,\theta}(r_i,\theta_{{\rm FL},i})=\partial_\theta f_i(\theta_{{\rm FL},i})\ .
\ee
It is somewhat complicated to calculate the radial derivative of flux function $\Psi_{,r}$
on a specific location $(r_i,\theta_{{\rm FL},i})$ along a field line.
Let us define a new vector of flux function $\Psi|_{\theta_{{\rm FL},i}}(r_i)$,
with $\theta\equiv\theta_{{\rm FL},i}$ and $r=r_i, i=1,2,\cdots$
The values of the new vector is obtained by 2-dimensional cubic spline interpolation
from the array $\Psi(r_i,\theta_{{\rm grid},j})$.
Then we fit the new vector with a polynomial function $h_i$ as
\be
\begin{aligned}	
	\Psi|_{\theta_{{\rm FL},i}}(r_i)&=h_i(r)\ ,\ i=1,2,\cdots\ ,
\end{aligned}
\ee
Thus, the radial derivative of flux function has the value
\be\label{eq:dPsidr}
	\Psi_{,r}(r_i,\theta_{{\rm FL},i})=\partial_r h_i(r_i)\ .
\ee

Furthermore, we refine the grids in the regions around $r_{\rm FM}^{\rm in}$ and $r_{\rm FM}^{\rm out}$.
Therefore, the two eigenvalues $({\mathcal E}/{\mathcal L})_{\rm in}^{(l)}$
and $({\mathcal E}/{\mathcal L})_{\rm out}^{(l)}$ can be well specified.

\item[\it Step 3] We obtain the global velocity $u^{(l)}$ of fluid from Step 2.
The terms $\mathcal{S}_{\rm EM}$ and $\mathcal{S}_{\rm MT}$ are updated by the four-velocity,
then fed into the MHD GS equation (\ref{GS_MHD}) for the next iteration.
Step 1 to Step 3 are iterated until all quantities converge to an expected precision.
\end{itemize}

Similar to what has been mentioned in Paper I, we need to smooth the density $\rho$ inside the loading zone where $u_p\equiv0$, since it is proportional to $\eta_m/u_p$ and therefore diverges.
To overcome this infinity, special physics inside the plasma loading zone need to be investigated in the future.

\subsection{Super-magnetosonic Outflow}\label{sec:collimj}

\begin{figure}[h!]
\centering
\includegraphics[scale=0.7]{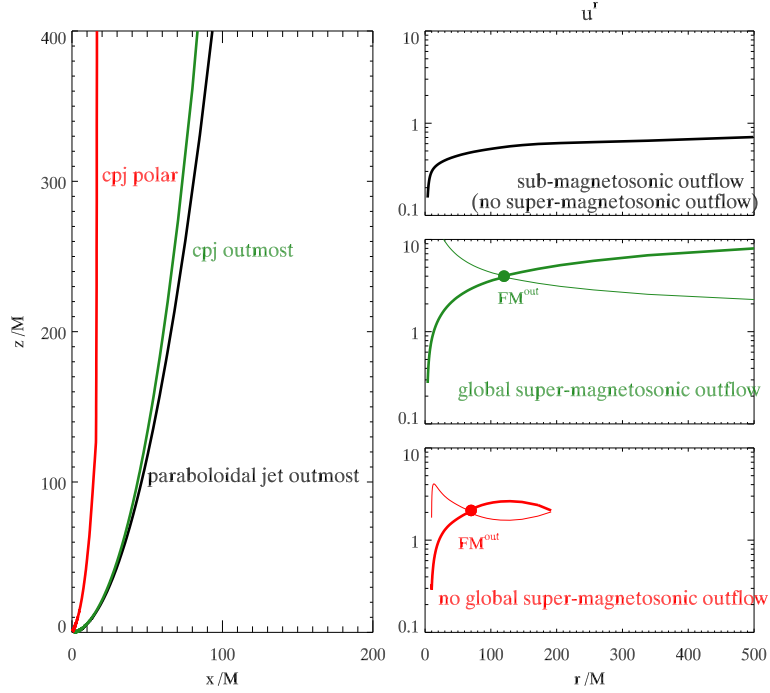}
\caption{ {\it Left:} Three examples of the poloidal configuration of field lines.
The black line represents the outmost field line of a paraboloidal jet.
The green line and red line represent a `cpj outmost' field line
and a `cpj polar' field line, respectively.
{\it Right:} The corresponding velocity of outflows related to the three examples.
Only sub-magnetosonic outflow exists ({\it Top-right}).
For the cpj outmost field line ({\it Middle-right}), the outflow passes FM point and becomes super-magnetosonic.
For the cpj polar field line ({\it Bottom-right}), there is no global super-magnetosonic solution in the computation domain.
\label{fig:collim}}
\end{figure}

With a specific choice of $({\mathcal E}/{\mathcal L})_{\rm in}$, the physical super-magnetosonic inflow
passes along $r_*\to r_{\rm A}^{\rm in}\to r_{\rm FM}^{\rm in}\to r_{\rm H}$.
The character of super-magnetosonic of the outflow, however, is sensitive to the poloidal shape of field line.

First, to be representative, let us consider the outmost field line of a jet in pure paraboloidal shape described by
\be\label{eq:parab}
	1=\frac{r+r_0}{r_{\rm H}+r_0}(1-\cos\theta_{\rm outmost}) \\
\ee
\citep{Tchek10,Nathan14,Mahlmann18}.
The poloidal configuration of this paraboloidal outmost field line with $r_0=10$ is presented in black line
in the left panel of Figure~\ref{fig:collim}.
The corresponding velocity $u^r$ of outflow is shown in the top-right panel.
We have found in practice that for any field line in the shape of pure parabole,
all physical solutions of outflows are sub-magnetosonic.
Similarly, only sub-magnetosonic outflow solutions exist for monopole field lines (Paper I).
The conserved quantity $({\mathcal E}/{\mathcal L})_{\rm out}$ must be determined by additional condition,
then the outflow passes along $r_*\to r_{\rm A}^{\rm out}\to\infty$.

Next, let us consider a collimated paraboloidal jet (cpj), with magnetic flux function described by Equation~(\ref{eq:collimj}),
$r_0=10$ and $\varepsilon_{\rm vt}=0.01$.
This is the initial guess of the fiducial model and is representative.
The poloidal configuration of the outmost field line is presented in green line in the left panel.
In the middle-right panel, we show two solutions of Bernoulli equation
with a correct eigenvalue $({\mathcal E}/{\mathcal L})_{\rm out}$ in thin green lines,
forming a standard `X'-type FM singularity. The thick line shows the physical super-magnetosonic solution which passes along
$r_*\to r_{\rm A}^{\rm out}\to r_{\rm FM}^{\rm out}\to \infty$.
It should be noticed that the velocity field is very sensitive to the field line configuration.
In the left panel, the red line represents a `cpj' field line in the polar region.
As shown in the bottom-right panel, there is no global super-magnetosonic solution in the computation domain adopted in this paper.
The corresponding outflow passes the outer FM point, but decelerates then breaks at $\sim200M$.
Therefore, with collimated field line configuration adopted,
the outer FM surface terminates on some field line close to the polar direction.
Beyond this field line there exists no global solution of super-magnetosonic outflow.
Fortunately, we find the fraction of such field lines is small.
Most field lines can carry physical super-magnetosonic outflows, if the value of $\varepsilon_{\rm vt}$ is not too large.

\subsection{Null Surface}\label{sec:null}

\begin{figure}[h!]
\centering
\includegraphics[scale=0.7]{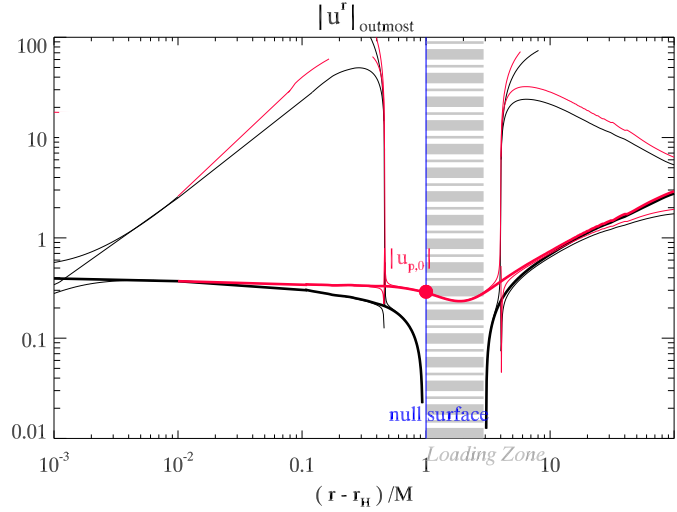}
\caption{ Solutions of contravariant radial velocity $u^r$ from Bernoulli equation for the outmost field line of our fidicial model. The black lines and red lines represent solutions with zero initial velocity and a specific value of $|u_{p,0}|$, respectively. The thick dashed gray line shows the loading zone. The thin blue line shows the null surface and the filled red circle shows the location of particle injection with non-zero velocity.
\label{fig:upst}}
\end{figure}

In this work, we assume that the pair production takes place at the null surface, where the current charge density becomes zero.
Assuming the initial velocity $u_{p,0}$ with which the plasmas are injected on a specific field line,
the total energy per particle $E$ can be calculated by
$({\mathcal E}/\eta_m)^2=(u_{p,0}+1)/U_g(u_{p,0};\ r_0,\theta)$.
Then the global solution of $u^r$ can be determined by searching the inner and outer FM surfaces.

We adopt the outmost field line of our fidicial model as an example.
We calculate two sets of solutions of $u^r$ with zero initial velocity ($|u_{p,0}|=0$)
and the non-zero one ($|u_{p,0}|>0$) for comparison.
In Figure~\ref{fig:upst}, we show all the solutions from the Bernoulli equation in black and red lines
for the former and latter set, respectively.
The physical solutions in both sets are thickened.
With zero initial velocity, there grows a special region where no solution of $u^r$ exits.
We call this region `{\it Loading Zone}', as shown in the thick dashed gray line.
$u^r$ equals to zero on both the boundaries, where the left is the null surface (blue).
Given a non-zero initial velocity, i.e., $|u_{p,0}|>0$ and $u_{p,0}^{\rm in}=-u_{p,0}^{\rm out}$,
the solutions of $u^r$ (red) exit inside the loading zone.
It can be found that the physical solution coincides the one with $|u_{p,0}|=0$ outside the loading zone.
Therefore, the assumptions $u_{p,0}=0$ and $u_p\equiv0$ inside the loading zone are simple but appropriate.
Neither the velocity with which fluids fall into BH, nor the velocity of outflow at large distances,
which we are most interested in, can be affected by $u_{p,0}$.
However, the physics in the loading zone remains as an open quesntion which should be studied in the future.

\end{document}